\documentclass[11pt]{article}

\usepackage[final]{acl}

\usepackage{times}
\usepackage{latexsym}

\usepackage[T1]{fontenc}

\usepackage[utf8]{inputenc}

\usepackage{microtype}
\usepackage{amssymb}

\usepackage{inconsolata}

\usepackage{graphicx}

\usepackage{booktabs}
\usepackage{array}

\title{LLMs in the Real World: Evaluating ``AI'' in Emergency Contexts}

\author{Sara Court \\
The Ohio State University \\
  {\tt court.22@osu.edu} \\\And
  Lara Downing \\
  Community Refugee \& \\ 
  Immigration Services (CRIS) \\
  {\tt ldowning@cris-ohio.org} \\\And
  Micha Elsner \\
  The Ohio State University \\
  {\tt elsner.14@osu.edu}}

\begin{document}

\maketitle

\begin{abstract}
This paper offers a call to action. We urge our colleagues in the research community to play a greater role in the articulation of our findings to the public. To illustrate the stakes we present a case study on the initial stages of an LLM-based machine translation application's deployment in a real-world context: a text-2-911 system advertising capabilities in 55 languages for use in emergencies in which it may be difficult to call operators directly. We identify a number of common misconceptions about technologies such as these, concluding with a set of concrete recommendations and best practices for stakeholders at every stage of the development and deployment pipeline. While the advancement of scientific research often lies in solving the ``hard'' problems, we argue it is often the ``easy'' ones--- problems for which the latest technology is often unnecessary--- that are most overlooked.
\end{abstract}

\section{Introduction}
\label{sec:intro}

Despite considerable overlap between academic and industry-based developers of Large Language Models (LLMs) and related technologies \cite{abdalla-etal-2023-elephant}, it seems Natural Language Processing (NLP) researchers have a science outreach problem. As our research findings continue to drive the development of some of the most quickly adopted user-facing applications to date \cite{debrugger2023}, the findings themselves--- and their real-world implications--- are too often lost in the hype. Artificial Intelligence (AI) is increasingly being offered as an inevitable solution to some of humanity's largest problems \cite{eubanks2018automating, ByrumBenjamin2022, Benjamin2024, cdt2025humans}.

As the outputs of modern NLP research are taken up and applied in commercial products, an information gap has developed between those designing NLP applications and their end users. For example, researchers may take it for granted that a model performs worse with lower-resourced languages \cite{silva2024benchmarking}, or that its performance degrades in response to inputs from outside its training domain \cite{wu2024retrieval, li-etal-2025-leveraging}. However, decision makers deploying NLP products in essential services like law enforcement and emergency response may not be aware of these limitations. Many may find themselves under pressure to find ways to acquire and integrate AI tools, but are left to navigate the AI software market without the knowledge needed to properly evaluate them, mitigate their risks, and ensure that they're as safe, ethical, and effective as possible.

This failure to effectively communicate our research findings to the public--- including to those developing and selling consumer-facing applications--- is not unique to NLP. Cryptographers have developed easy-to-access, publicly available code for secure public-key communication, but surveys of these systems in actual use persistently reveal that end users continue to create security flaws by misusing the APIs \citep{choudhari-cryscanner,lazar-fail}.
The NLP research community now faces a similar problem. 
Although basic tools for transparency and evaluation like model cards \citep{mitchell2019model} have existed for years, they still aren't widely used in commercial settings. 

Most consumers of NLP technologies have little other than promotional materials to go off of as they navigate the complex landscape of tools marketed to the public as ``AI.'' The result, we claim, is both an information imbalance and an accountability gap: NLP products are increasingly being sold to consumers in both public and private sectors, including in high stakes contexts such as policing \cite{USvCruzZamora_2018, quaglia2022death}, immigration court \cite{Deck2023}, critical public health announcements \cite{Moreno2021}, and other emergency response \cite{Burns2025}, without adequate information or support to use them safely. Furthermore, if harm occurs as a result of technological error or misuse, it is often unclear who, if anyone, can be held accountable for that harm. We believe this should be cause for genuine concern within the research community. When language technology is deployed to support emergency services within our own communities, the stakes are high and all of us are stakeholders.

This paper presents a case study on one example of an LLM-based language technology already deployed for use in emergencies at a local 911 center in the United States. 
We describe the technology's rollout via its marketing and promotional materials, and describe our experience meeting with staff involved in its deployment at the 911 center.

Our experience sheds light on a number of common misconceptions about language and language technology that, in combination with systemic gaps in accountability, can
result in the risky and potentially harmful deployment of NLP systems.
This is of particular concern in situations such as our case study, in which the product is being used in emergencies and affects some of the most vulnerable members of our community--- refugees and other immigrants for whom English is not a native language.
We discuss the role that NLP researchers
can play in addressing these critical gaps, and conclude the paper with a set of concrete recommendations for best practices.
Finally, we encourage our colleagues in the research community to do more to support science outreach,
so that our advice may be audible to those who need to hear it.

\section{Case Study: Text-2-911 Service}
\label{sec:casestudy}

\subsection{Language Access and Technology in Emergencies}
\label{sec:background}
According to U.S. federal and state statutes, including Title IV of the Civil Rights Act of 1964, the Americans with Disabilities Act, the Affordable Care Act, and the 14th Amendment, among others, emergency service providers are legally obligated to ensure language access for callers with limited English proficiency. While the ability to communicate with emergency responders may be assumed as a given if one's native language is English, lack of equitable language access can compound hardships faced by many of our most vulnerable populations, including immigrant and refugee communities, as well as individuals with disabilities \cite{NiWAP2013LanguageAccess, taira2021pragmatic, bhuiyan2023lost, hofmann2024ai, parmar2025axon}. Community organizations can provide crucial support--- sometimes the only support--- for such individuals, helping them to navigate complex systems and institutions with which they may be unfamiliar, educating them about their rights, and connecting them to critical resources. 

The second author of this study is a licensed social worker at one such community organization, leading a multilingual team of victim advocates specialized in serving immigrant and refugee survivors with limited English proficiency. Through this work, she and her colleagues have seen firsthand how a lack of qualified interpreters and misuse of language technology can lead to a cascade of harmful effects on vulnerable populations. So when the city announced the rollout of a new AI-powered\footnote{Despite repeated attempts to determine the exact model architecture powering the service, we are still not 100\% certain of its design.} translator for text-2-911 emergency response, she and her team were eager to learn more.

The option to send text messages via SMS directly to 911 emergency call takers in English began as a solution for Deaf and Hard of Hearing callers and
``those who may be unable to communicate verbally due to background noise or safety considerations'' \cite{laird2025} and became available in early 2019. Crucially, the text-2-911 tool is not intended to be an equivalent alternative to a voice call \cite{franklintext2911}. 
In our visits to the 911 center, described in Section \ref{sec:visiting}, the staff made clear that voice calls are always preferred since they can provide additional information, such as background noise and voice distress, to the call operator. Dispatchers are trained to ask anyone who texts the 911 center whether they are able to call safely, and encourage them to do so if possible.
The city already provides human interpreters for voice calls to serve the estimated 6.4\% of area residents over 5 years old who speak English ``less than very well'' \cite{cohc2025healthmap}, and interpreter services were used for about 4,000 of the 670,000 total calls to the local 911 call center in 2024 \cite{laird2025}.

\subsection{Popular Perceptions of Machine Translation}

More often than not, press coverage of MT applications portrays systems in an overwhelmingly positive light, with relatively little scrutiny of the limitations of the technology or its implications for human interpretation \cite{vieira2020understanding}. The deployment of MT systems, even in high stakes contexts, is often contrasted with the option of providing no language access rather than compared to the statutory baseline, namely in-person or teleinterpretation by qualified, human translators \cite{quaglia2022death}. When the bar is set artificially low, it is easier for well-intentioned community members to celebrate the use of MT as ``above and beyond'' when it may actually represent a step backward in access for callers with limited English proficiency. 

In our case, one local media report suggests that fear of language barriers when calling 911 ``may soon be a thing of the past.''
The article goes on to state that ``instead of relying on language interpreters to help non-English speaking callers [...] callers can now text 911 in their own language,'' contradicting the stated intent of the service not to supplant voice-call interpretation but rather to add an additional accessibility option \cite{keller2025columbus}. 

City residents interviewed about the new technology for the promotional video echo the language from the original press release, assuring residents that they could now text 911 ``in their native language.'' However, a list of the 55 languages supported by the model has not been
included in any press coverage, and we were only able to acquire it by contacting the 911 center directly. Not all of the county's most commonly spoken languages are on the list, and non-Latin scripts can only be sent through AT\&T, a disclaimer that could be easily missed in much of the press coverage.

It was clear that more information was needed if the second author's victim services team wished to provide accurate information and guidance to their clients. This is when the second author reached out to a local university's linguistics department for clarity on current MT technology. She also contacted the 911 center staff, who immediately and graciously invited her team to visit the center for a tour and in-depth discussion of the new features. Due to ongoing updates to the software, they were not able to test out the translation on that day, so a second meeting was arranged, and the first author was invited to join. The following section describes what was learned from these meetings. 

\subsection{Visiting the 911 Center}
\label{sec:visiting}

The first and second authors visited the 911 center at the end of September 2025, along with two colleagues from the victim services program who were eager to test out the translation tool in their respective native languages.
Each had prepared a list of phrases from real-world text messages to explore how the model handled language-specific challenges such as dialect variation, text speak, typos, referential ambiguity, idioms, and code switching. Three city staff members from the 911 center and a representative from the software company providing the MT application generously made time to host us, even as they orchestrated the day’s emergency response activities for a city of 900,000 residents. 

The 911 center is home to the city’s Public Safety Answering Point (PSAP), where operators receive all incoming local calls and texts to 911 before routing them to the appropriate responders, e.g., fire, EMS, or police. 
Software and maintenance for the PSAP interface, including the text-2-911 feature, is provided by a third party who advertises their use of Microsoft Azure to provide language detection and automatic translation. 
According to 911 center staff, Microsoft does not provide access to the underlying model or training data.

The staff managing implementation of the tool had also not been provided any evaluation data or quality assurance services by their software provider.
While a policy exists at the state level outlining deliberate and detailed requirements for ``planning, implementation, procurement,
security, privacy, and governance requirements for the use of Artificial Intelligence (AI)'' \cite{OhioIT17_2023}, no equivalent policy has been created for City departments. This appears to leave the 911 center without the necessary subject-matter expertise, training, guidance, or resource allocation to ensure that proper safeguards are in place. 

According to the software company's representative, the goal of the translation tool is to decrease response time for end users with limited English proficiency. However, there is currently no ongoing evaluation to establish the product's success toward meeting that goal. 
The MT system also does not integrate any oversight from human translators, either in real-time or after-the-fact for quality assurance. A human dispatcher receives and responds to the translated text, but the text output by the MT model, as with any AI model, is still ultimately AI-generated.

\subsection{Testing the Tool}

We were given the opportunity to interact with the system ourselves in real time. We probed the model with common linguistic phenomena found in text messages, such as accidental misspellings and dialectal variation. 
Ultimately, both of our colleagues from the Victim Services Program encountered challenges texting 911 in their native languages, Arabic and Nepali, respectively. 

In the case of Arabic, we learned that Modern Standard Arabic (MSA) was the only variety of Arabic in which the MT system was supposed to be able to interact \cite{al2025evaluating, mishra2025investigating}. However, in addition to the limitations posed by Arabic’s non-Latin orthography, those familiar with the sociolinguistic contexts in which Arabic is spoken will know that MSA is rarely, if ever, the language a speaker will use to communicate via text message. Dialectal variations in lexical items and spelling presented clear challenges for the model's ability to interact with an Arabic speaker via text.

Similarly, the MT system is only able to recognize Nepali written in the Devanagari script. However, at least among the Bhutanese-Nepali community making up the majority of Nepali speakers in the area, text messages are almost exclusively written using the Latin script. Our colleague was not even sure how to use a Devanagari keyboard, and could not find all of the symbols he would need to interact with the system using the orthography that the model was trained on for the language.

The potential negative impact of such
incongruence between the data the model was trained on and that which might be encountered in a realistic setting is further amplified by 
the tool's lack of an informed consent procedure for end users. When a person tries to contact 911 via text message in a language other than English, the interface displays both source text and translation output to the dispatcher at the 911 center, who typically is not proficient in the language being translated. In contrast, the use
of MT is not explicitly disclosed and neither the translations nor the name of the language detected by the model 
are presented to the person texting 911. On the surface, their experience is no different than if they were texting directly to another person.
They receive no disclosure of AI-generated translation or user guidance for optimizing output accuracy. 

There is a reasonable concern that wordy disclaimers or excessive instruction could slow down the response time or cause confusion in an emergency situation. These are valid considerations that warrant empirical investigation. However, lacking that evidence, and given the vast amount of evidence of the errors and risks associated with LLM technologies already documented by the research community \cite{costa2015linguistically, berk2021artificial, freitag2021experts, mehandru-etal-2023-physician, court2024shortcomings, freitag-etal-2024-llms, mickus-etal-2024-semeval, urlana2025no}, we wish to emphasize that requiring informed consent and transparency for users on both sides of the interaction is not only more ethical, it is also likely to improve the tool's overall performance and make the service more effective.

During the meeting at the 911 center, we discussed a number of additional ethical considerations and safety precautions. Although most of our concerns have been well-documented in the academic literature for years (see, e.g., \citeauthor{kumar2023language} \citeyear{kumar2023language} for an overview), public officials and local decision makers are often unaware of many of the ethical best practices NLP researchers might now take for granted \cite{karamolegkou-etal-2025-ethical}. There is still a general lack of understanding of the potential vulnerabilities and risks inherent to these technologies, reflecting what we see as an overall gap in access to information and insufficient involvement or support from experts in our field.
Addressing these issues at all stages of the model's life cycle likely requires action by city leadership, including legislative policy, resource allocation, and partnerships with local community organizations and members of the NLP research community.

\subsection{Learning from Experience}
We hope our case study will encourage our colleagues in the research community to play a greater role in advocating for the safe and responsible application of their own findings. 
However, it should be noted that although a sizeable number of ACL submissions each year are authored or co-authored by researchers in the private sector \cite{abdalla-etal-2023-elephant},
NLP research culture itself has continued to shift away from open-source principles and peer-reviewed science towards greater secrecy and a ``move fast and break things'' approach
that prioritizes 
profits
and tends to benefit only a small subset of the global population \cite{benjamin2019race,Blodgett_Barocas_III_Wallach_2020, junker2024data}.

While the rapid adoption of MT has raised concerns across domains, significantly more attention and resources have contributed to a greater body of research and evidence-based approaches to MT integration in fields such as emergency medicine \cite{Dew2018,Lopez2025,Anyaegbuna2026}. In contrast, many of the groups buying and selling these technologies in other domains lack the technical expertise, clear guidance, or necessary resources to audit and evaluate the deployment at the necessary scale \cite{vieira2020understanding}. In the multitude of situations in which a researcher isn't present to evaluate an AI product with a critical eye, the success of an LLM application's deployment depends heavily on the pre-existing knowledge and abilities of those acquiring and using it.

The following section describes a number of specific misconceptions about language and language technologies that we've repeatedly observed in circulation among the general public.
Perpetuated by the broader societal patterns described in Section \ref{sec:persists} and without the support of experts in our field to counteract them, we believe these gaps in knowledge will continue to enable instances of inappropriate, ineffective, and sometimes even harmful deployment of LLM-based language technologies.

\section{Common Misconceptions about ``AI''}
\label{sec:misconceptions}

Computer science knowledge or AI literacy among those buying, selling, using, and regulating NLP technologies has consistently lagged behind the speed at which the field has advanced. Journalists who may otherwise provide a source of oversight and information are often also un- or under-informed, which can mask important considerations and mislead the general public \cite{vieira2020machine}. It is worth asking where the following misconceptions come from, and we encourage the research community to do more to publicly debunk them.

\subsection{Misconception 1: The Term ``AI'' is Well-Defined}

Since its inception as a field of study, there has been debate about what actually constitutes artificial intelligence \citep{turing1950}. 
``AI'' has become a catchall term for a wide variety of large pretrained models, many of which are rapidly becoming a part of our everyday landscape. This can generate an unfortunate--- and inaccurate--- impression of homogeneity, obfuscating the difference between NLP tasks and objectives, such as those involved in machine translation vs. a dialogue system, or between model architectures, such as the distinctions between LLM-based pipelines and traditional NMT.
Conflating these systems into the umbrella term ``AI'' contributes to the belief that all of this technology is the same, with the same capabilities, costs, and problems.

\subsection{Misconception 2: AI has Superhuman Intelligence}
\label{sec:superhuman}

It is common for even researchers to anthropomorphize language technologies that sometimes display what can feel like superhuman abilities, like recalling specific facts about more than an encyclopedia's worth of topics \cite{deshpande2023anthropomorphization, Erscoi_Kleinherenbrink_Guest_2023}.
Having already passed the Turing test with flying colors for years, models may now be marketed to consumers as possessing or approaching Artificial General Intelligence (AGI)--- a markedly superhuman ability whose actual definition is just as vague and debatable as any other kind of intelligence \cite{mahowald2024dissociating, mitchell2024metaphors}.
Assuming there is no viable alternative,
it is understandable that consumers looking to serve speakers of languages other than English might turn to a seemingly superhuman MT system in an effort to provide something rather than nothing.
Even with the best of intentions, however, confusion between supporting a language and supporting it well can have dire consequences \cite{bhuiyan2023lost, calmatters2025deaf, cdt2025content, quaglia2022death, Deck2023}.

\subsection{Misconception 3: Language is Easy}

Potential users of ``AI'' are not just unaware of the fine points of language technology; they often also hold a variety of misconceptions about language itself \cite{wagner2023extent}. These issues can be mutually reinforcing--- some have told us they want ``translation'' rather than ``interpretation'' because they want to know word-for-word what their interlocutor is saying. Linguists and translation theorists know that this isn't the right approach: the individual words don't always communicate the core meaning, and utterances can be ambiguous or multivalent even for an experienced interpreter \cite{nielsen2025alligators}. But this folk theory of translation contributes to the misguided belief that machine translation is more objective and therefore more accurate than any human interpreter. 

People may also hold misconceptions about language diversity, for example assuming that dialectal variation is merely a matter of accent or that stigmatized dialects are language errors resulting from poor education \cite{hudley2024decolonizing}.
Given such misunderstandings, it may be easy to believe that a technology advertising support for over 50 languages will be able to serve all of them equally and that dialectal variation will not cause significant problems, contrary to findings from NLP research \cite{aycock2024can, hofmann2024ai}.

\subsection{Misconception 4: Quantitative Metrics are Reliable and Sufficient}

As a largely empirical discipline, NLP relies heavily on automatic and quantitative metrics. NLP researchers commonly acknowledge the inadequacies of their own metrics \cite{flamich2025birds} and may even take part in shared tasks attempting to improve them \cite{freitag-etal-2024-llms, shayegh2025feeding}. Unfortunately, awareness of a metric's limitations too often fails to make it beyond academic circles. 
In contrast, techniques used to market and sell LLM technologies leverage benchmarks to advertise some of the ``superhuman'' capabilities discussed in Section \ref{sec:superhuman}.
End users may be unaware that even the best performing model will degrade outside its training domain \cite{saunders2022domain}, and benchmark scores can be gamed \cite{mansurov-etal-2025-data}. 
Moreover, simply interpreting the metric numbers can be difficult for novices. Long experience of evaluation gives professionals a general notion of how to mentally map between MT metric scores and translation quality \citep[e.g.][]{scarton-etal-2019-estimating}.
Without this experience, 
one might incorrectly assume that a high score means that mission-critical errors have already been eliminated.

\subsection{Misconception 5: Technological Solutionism}
\label{sec:solutionism}

The problem of over-estimating the abilities of technology while underestimating our own is not a new one. 
``Solutionism'' \cite{morozov13} is the tendency to assume that social problems are amenable to engineering solutions--- especially quick, cheap and disruptive ones. The misconception is not that technology cannot help; it often can! But in order to do so, it needs to be embedded within a supportive social context \cite{Benjamin2024, sanchez2025ethical}. Many of the problems AI is being sold to fix would likely be more efficiently and effectively resolved with simpler methods \cite{quaglia2022death}.
Instead, vendors of so-called ``AI solutions'' often market their products by communicating, directly or indirectly, that the human element can be dispensed with entirely. 

However, human oversight is essential when deploying technologies as erratic, unpredictable, and potentially even deceptive as LLMs \cite{ouyang2022training, Roose2023BingChatbotTranscript, mickus-etal-2024-semeval, cdt2025humans}.
LLM software providers should therefore be expected to provide training and support for human quality assurance teams, as well as ongoing monitoring in collaboration with professional human interpreters to systematically collect and review feedback from the app's end users.
The question ``what if something goes wrong'' is central to engineering robust systems \citep[ch. 1.6, ch. 10]{kapur-reliability}. Failure to ask and answer this question can make a system appear relatively cheap to deploy, but this is only because its true costs appear primarily in scenarios where it \textit{doesn't} work. 
The reported cost of responding to a domestic violence homicide, for example, stands in the millions of dollars \cite{VonNessen2025ODVN}. 

Such errors are not inevitable, and their related losses could be minimized by prioritizing
the training and employment of professional human interpreters over high-tech solutions, particularly when it is not possible to guarantee the safety of the technology in question. The results of doing so would not only be ``better than nothing,'' they would be quantifiably better than a machine translation system of variable or unknown quality. Investment in humans and the things we do best, such as language and translation, would thus make for a sound financial (as well as ethical) decision.

\section{Why the Problem Persists}
\label{sec:persists}

We believe our case study represents broader trends
in the deployment of LLM-based language technologies around the globe.
Why does this happen?

\subsection{Information Asymmetries}

Without a doubt, one of the biggest contributing factors to the inappropriate and sometimes unethical use of LLM-based technologies
is that consumers, even with the best of intentions, lack access to information.
There is an AI literacy crisis at nearly every level of the adoption chain.

Information asymmetries begin before many pretrained models are even released to the public.
In many cases, it is only possible to infer what the model was trained on by considering its outputs in light of the old computer science adage: ``garbage in, garbage out.'' 
The details of a model's development are further obscured once it is packaged into software and sold by a third party.
Those selling the software do not necessarily understand how their product works in technical terms, and even when developers understand the API they are using, they have not necessarily been trained in the core technologies behind it.

Analogous to the problem faced by cryptographers described in Section \ref{sec:intro}, the mass availability of APIs for LLMs creates the illusion that no specialized knowledge is needed to use the product. Similar to using a database server or hash function, it's easy for an engineer to assume that a large company like Microsoft or Meta has made their product available because it ``works,'' without a full understanding of what ``working AI software'' means or the inherent risks and potential errors that come with deciding to use it.

The research community is not without our share of responsibility, either. Although we have made much progress towards an agreed-upon set of standards for conducting ethical research, the economic and cultural environments in which this work takes place do not tend to value or support science outreach and communication to the public.
Regulators and legislators also often lack AI literacy, and their 
perception of NLP research is dominated by the perspectives of a small handful of powerful companies \cite{nyt2025trump}. Without the infrastructure and support to quickly and directly communicate our findings to the public, academic researchers effectively surrender our ability to speak on behalf of our own science.

\subsection{The Accountability Gap}

For NLP software development to be both safe and effective, knowledge has to move from the research community through multiple layers of transmission. At each of these layers, there is an ``accountability gap'' to cross: developers and sales people won't learn best practices unless they have a good reason to \cite{eubanks2018automating, Hohenstein_Jung_2020}. 
Existing regulations offer only partial coverage in response and tend to be geographically fragmented. For example, both the EU AI Act \cite{euaiact2024} and the Colorado AI Act in the U.S. \cite{coloradoaiact2024} distinguish high-risk AI systems and require additional transparency, oversight, evaluation, and other obligations. 

Policy and guidance at the federal level in the U.S. has been fragmented and slow to adapt. For example the NIST AI Risk Management Framework \cite{nist_ai_rmf_2023} omits specific provisions for language access and generally leaves enforcement mechanisms up to voluntary self-regulation. 
In April 2026, NIST announced the launch of a new ``Profile on Trustworthy AI in Critical Infrastructure'' \cite{nist_cip_2026}. While details remain forthcoming at the time of writing, we note that this action comes nearly three years after the original risk management framework was published.

Some domains, for example medicine and law, already have well-established systems of individual and institutional accountability that can be applied to NLP tools, for example by transparently defining regulations, formalizing community norms, and creating community-internal resources for learning about these technologies \cite{vasey2022decide, landers2023auditing, lekadir2025future}. In policing and emergency response, the situation seems far less structured \cite{taira2021pragmatic, parmar2025axon}.
Without independent, impartial evaluation based on the technical details of the model and its specific use context, even well-intentioned actors are left without adequate guidance on how to use the technology safely.

\section{Recommendations and Best Practices}
\label{sec:recs}

We wish to reiterate that the text-2-911 service and our experience at the 911 center reflect broader patterns in AI software deployment rather than an edge case. Similar accountability failures have been observed across other high-stakes contexts \cite{mahase2023babylon,italydpa2025, moseretal2025}. In each example, we see a common thread: NLP technologies are being deployed in our communities without adequate support to properly evaluate their performance and limitations, and without clear mechanisms in place to identify and mitigate their potential risks and harms. In the end, not every problem calls for a high-tech solution, and not every technology is actually capable of solving the problems it claims to address. When AI is positioned as a universal solution to language access in critical contexts, the very tool intended to expand language access may ultimately reduce it instead.

\subsection{Deploying Language Technologies in High-Stakes Situations}

In an alternative scenario, staff at the call center in our case study would have had a much clearer idea of what sort of product they were getting. This should begin at the point of sale: if a company advertises translation in multiple languages, they should be transparent about the relative performance of each language pair and forthcoming about the potential risks and limitations of their software. For the system we examined, this means clearly communicating differences in translation accuracy across language pairs, specifically under conditions representative of emergency situations, using both quantitative (automatic) \textit{and} qualitative (human) metrics for evaluation.

Just as it is no longer acceptable to buy packaged food without a nutrition label or drugs without a pharmacist's consultation, model cards \cite{mitchell2019model} should be required for all software applications trained using machine learning methods. Similar to a pharmaceutical, model cards should distinguish between on-label and off-label uses and clearly communicate the known, potential, and hypothetical risks of deployment in the particular contexts for which they are intended. 

An explicit analysis of model failures should also be an expected part of deciding whether to acquire a new language technology prior to its deployment. This would allow organizations to formulate a contingency plan for any errors they observe in the process or otherwise believe to be probable. Ideally, this would also allow for a more informed and efficient use of public resources, including spending on qualified human interpreters to provide backup for the most heavily used language pairs and sensitive applications. Organizations can make these decisions more responsibly by formulating clear policies and standards before acquiring or using any specific product. For suggested minimal policy recommendations to specifically address language access when using AI technologies, we refer the reader to the SAFE-AI Task Force Guidance (\citeyear{safeai2024guidance}, \citeyear{SAFEAITF_2025_PartA}), which we adapt and present in Table \ref{tab:safeai}.

\begin{table*}[htbp]
\centering
\small
\begin{tabular}{p{0.92\textwidth}}
\toprule
\textbf{Suggested Policies for Ethical Use of AI for Interpreting (Minimum Requirements)} \\
\midrule
$\checkmark$ \textbf{Informed Consent} to accept or decline the use of an AI product \\[0.3em]
$\checkmark$ \textbf{Opt In/Opt Out} of data collection and storage without penalty \\[0.3em]
$\checkmark$ \textbf{Ability to Shift} between AI and human interpreting at any point in a timely manner \\[0.3em]
$\checkmark$ \textbf{User-Friendly Grievance Process} to report errors or harm \\[0.3em]
$\checkmark$ \textbf{Clear Explanations} of policies regarding privacy, confidentiality, and degrees of AI involvement \\[0.3em]
$\checkmark$ \textbf{End-User Autonomy} throughout the interpreting process \\[0.3em]
$\checkmark$ \textbf{Evidence of Improvements} to end-user wellbeing and safety are made available to the public \\[0.3em]
$\checkmark$ \textbf{Public Transparency} of quality metrics and their results over time \\[0.3em]
$\checkmark$ \textbf{Accountability and Oversight} to hold providers responsible for any errors or harm \\
\bottomrule
\captionof{table}{Ethical principles for the use of AI for interpreting, adapted from \citet{safeai2024guidance}.}
\label{tab:safeai}
\end{tabular}
\end{table*}

In general, we recommend consumers pay more careful attention to matching their deployment context(s) with appropriate levels of technological maturity and reliability. While it might be tempting to reach for the most powerful models for applications serving our most critical use cases, we need to be setting the bar higher, not lower, in such scenarios. The idea is not that we shouldn't be using LLMs in emergencies, but rather that these models need to be more closely evaluated and monitored before deployment. Perhaps 911 isn't the best call service to pilot automatic translation. 

Once a system is deployed, it should be continually monitored and evaluated on the real data it faces and its scores should be publicly available.
Community partners can be valuable collaborators for
this kind of evaluation, since they are likely to be best informed of the actual needs and nuances specific to the populations being served. Their involvement may also help to distinguish the errors that really matter from those with less serious consequences, allowing emergency service providers to better allocate their limited resources. 

User interfaces and interactions with these technologies also need to be more immediately transparent. 
Both parties should be able to see how the system translates their messages and confirm that the correct language has been identified. End users are more likely to have enough English proficiency to identify errors than dispatchers with little or no exposure to the target language, as well as prior experience with MT performance in their native language. Increasing transparency allows for more robust informed consent in real time.

While appropriate AI governance requires funding beyond the product itself, such expenses are negligible in comparison to the hidden costs, monetary or otherwise, of bypassing human oversight and getting it wrong when it really matters.

\subsection{How the ACL Community Can Help}

As researchers, we should remind ourselves periodically that the complex problems we may be trying to solve are not necessarily the biggest issues that stakeholders still face. Model cards and help lines may not seem cutting-edge, but many of us could also use a reality check: NLP practitioners sometimes over-estimate what's considered common knowledge or how much most people actually understand about language technologies. 

We also encourage more active contributions to local AI literacy initiatives. That is not to say that every researcher ought to also do science outreach, but that we can all strive to be better about supporting those among us who do. This can take the form of financial support, but may also just mean advocating for these colleagues within our networks. For example, the ACL could recognize researchers involved in public outreach and safety, as some other professional organizations do (e.g., the Linguistics, Language, and the Public Award presented annually by \citeauthor{lsa_public_award}).

Finally, we ought to consider coming to a public consensus on key terminology in our field. NLP researchers know that AI is not one thing, so it's important to communicate this to the public.
We can advocate for the use of more specific terminology when describing these products--- for example ``LLM chatbot'' or ``machine-generated translation''--- which may help those outside the research community better distinguish between the various models and their intended uses.

\section{Conclusion}
\label{sec:conclusion}

This paper presents a case study on a situation we believe to be representative of wider patterns in language technology deployment with the potential to inflict serious, but preventable, harm.
When accurate translation can mean life or death, the technology providing it needs to be deployed as ethically and safely as possible.
Regulatory legislation and community education are both important, but without active engagement from the research community these strategies are unlikely to be sufficient to address the range of issues we've described. There must also be clear mechanisms in place to hold accountable those developing, selling, and providing these technologies once NLP research has moved beyond the theoretical or academic realm. In our opinion, it is both unethical and impractical to place the burden of responsibility on the consumer when deploying LLMs in such critical situations as the one we've described in this study.

Our intention in sharing these experiences is not simply to criticize developers or users of NLP applications.
Rather, we wish to facilitate open discussion among members of the NLP research community, and across the entire web of stakeholders deploying the technologies our research supports. We hope this may be able to at least mitigate some of the harms that can result from our findings making their way into society, improving the quality of these applications and increasing their beneficial impact in order to make our communities safer for \textit{everyone}. 

\section{Limitations}
\label{sec:limitations}
The scope of the current paper is limited to one case study, but we believe the conclusions and recommendations we draw from it apply more broadly. We base our discussion primarily around the language used when advertising the MT tool described, as well as our meetings with call center staff, with whom we hope to continue collaborating in order to improve the quality and safety of the services being offered. Not only are so-called ``AI'' systems themselves relatively new innovations, the software we evaluate as part of our case study has only recently been deployed live. There have yet to be enough interactions with the service, nor have we been invited by the call center or software provider, to conduct the kinds of statistical analyses typically used to validate empirical research in NLP. As we hope to have made clear, describing the technology as ``AI-powered'' also limits our ability to know or describe the exact model(s) being used or how the system was designed. Finally, ethical considerations, described in the following section, also place rightful limits on the data and methods we might use to further evaluate the tool described in our study.

\section{Ethics Statement}
\label{sec:ethicsstatement}
This paper addresses life or death services for a segment of our community that is currently among the most vulnerable and the most targeted. In doing so, it was essential for attention to be brought to the risks resulting from a lack of AI literacy in the community, without needing to spotlight individual experiences. No data contained in this paper was obtained, directly, or indirectly, through the provision of services for victims of crime or funded with dollars intended to support those services.   

Care was also taken to avoid ascribing bad faith to any of the local stakeholders or to assign blame for the issues described in Section \ref{sec:persists}, all of which are far from unique to our case study. Every public official, first responder, and community member we spoke to has been open and committed to the goal of increasing language access. This is precisely why we feel an equal sense of responsibility to seek out the role that we, as individuals and as a field, can contribute to the realization of that goal.

\section*{Acknowledgments}
We are grateful to the call center staff who made the time to meet and talk with us at length about their adoption of this technology, and to the refugees, immigrants, and advocates whose personal comments on their experiences with language access helped define the social circumstances of the project. In particular, thank you to Anoj Sharma and Lina El-Zein, who tested the product on behalf of the communities they serve and explained the unique challenges they encountered in the use of Nepali and Arabic, respectively. We also thank three anonymous reviewers for their useful suggestions.


\bibliography{ custom2,anthology-1,anthology-2}
\clearpage

\end{document}